# Effects of spatial size, lattice doubling and source operators on the hadron spectrum with dynamical staggered quarks at $6/g^2 = 5.6$

Khalil M. Bitar, R. Edwards, U. M. Heller and A. D. Kennedy

*SCRI, Florida State University, Tallahassee, FL 32306-4052, USA*

Steven Gottlieb

*Department of Physics, Indiana University, Bloomington, IN 47405, USA*

J. B. Kogut

*Physics Department, University of Illinois,*
*1110 West Green Street, Urbana, IL 61801, USA*

A. Krasnitz

*IPS, RZ F 3, ETH-Zentrum*
*CH-8092 Zurich, SWITZERLAND*

W. Liu

*Thinking Machines Corporation, Cambridge, MA 02139, USA*

Michael C. Ogilvie

*Department of Physics, Washington University, St. Louis, MO 63130, USA*

R. L. Renken

*Physics Department, University of Central Florida, Orlando, FL 32816, USA*

D. K. Sinclair

*HEP Division, Argonne National Laboratory,*
*9700 South Cass Avenue, Argonne, IL 60439, USA*

R. L. Sugar

*Department of Physics, University of California, Santa Barbara, CA 93106, USA*

D. Toussaint

*Department of Physics, University of Arizona, Tucson, AZ 85721, USA*

K. C. Wang

*School of Physics, University of New South Wales, Kensington, NSW 2203, Australia*

# ABSTRACT

We have extended our previous study of the lattice QCD spectrum with 2 flavors of staggered dynamical quarks at $6/g^2 = 5.6$ and $am_q = 0.025$ and $0.01$ to larger lattices, with better statistics and with additional sources for the propagators. The additional sources allowed us to estimate the $\Delta$ mass and to measure the masses of all mesons whose operators are local in time. These mesons show good evidence for flavor symmetry restoration, except for the masses of the Goldstone and non-Goldstone pions. PCAC is observed in that $m_\pi^2 \propto m_q$, and $f_\pi$ is estimated. Use of undoubled lattices removes problems with the pion propagator found in our earlier work. Previously we found a large change in the nucleon mass at a quark mass of $am_q = 0.01$ when we increased the spatial size from 12 to 16. No such effect is observed at the larger quark mass, $am_q = 0.025$. Two kinds of wall source were used, and we have found difficulties in getting consistent results for the nucleon mass between the two sources.

# 1 INTRODUCTION

Calculations of hadron spectroscopy remain an important part of nonperturbative studies of QCD using lattice methods. (For reviews of recent progress in this field, see Ref. [1].) We have been engaged in an extended program of calculation of the masses and other parameters of the light hadrons in simulations that include the effects of two flavors of light dynamical quarks. These quarks are realized on the lattice as staggered fermions. We have carried out spectrum calculations with lattice valence quarks in both the staggered and Wilson formulations.

These simulations are performed on $16^3 \times 32$ lattices at lattice coupling $\beta = 6/g^2 = 5.6$ with two masses of dynamical staggered fermions, $am_q = 0.025$ and $am_q = 0.01$. These are the same parameter values as we used in our first round of simulations[2]. However, the first set of simulations had two known inadequacies. The first was that most of our runs were carried out on lattices of spatial size $12^3$. A short run on $16^4$ lattices with dynamical quark mass 0.01 showed that the $12^4$ lattices were too small: baryon masses fell by about fifteen per cent on the larger lattice compared to the smaller one. Thus, it was important to investigate finite size effects for $am_q = 0.025$. We also felt the need for more statistics on the $am_q = 0.01$ system for lattices of spatial size $16^3$.

Second, nearly all of our earlier running was done on lattices of size $12^4$ or $16^4$; these lattices were doubled (or quadrupled) in the temporal direction to $12^3 \times 24$ (or $12^3 \times 48$) or $16^3 \times 32$ for spectroscopy studies. Doubling the lattice introduced structure in the propagators of some of the particles: the pion effective mass, in particular, showed peculiar oscillatory behavior as a function of position on the lattice. This behavior was almost certainly due to doubling the lattice [3] and the best way to avoid this problem is to begin with a larger lattice in the temporal direction. Because of these difficulties, mass estimates from such doubled lattices are suspect. This is seen when comparing the masses obtained from the doubled or quadrupled $12^4$ lattice with those from the $12^3 \times 24$ lattice in our previous work.

In our work on smaller lattices, only one kind of source was used, the so-called "corner" source. In these simulations we include a second kind of source (in fact 3 sources) which enables us to measure the $\Delta$ mass, as well as the nucleon. Furthermore, with these new sources, we are able to measure masses of all mesons created by operators which are local in time, and correspond to strictly local continuum operators (local quark bilinears with no derivatives). This allows us to study the extent to which flavor symmetry, which is broken by the staggered lattice, is realized at this lattice spacing.

Some of the results described here have been presented in preliminary form in Ref. [4]. Studies with Wilson valence quarks which complement the results presented here have been published [5] as have studies of Coulomb gauge wave functions [6]. In addition, we are preparing a paper on glueballs and topology. In Sec. 2 we describe our simulations and in Sec. 3 we give our results and conclusions.

## 2  THE SIMULATIONS

Our simulations were performed on the Connection Machine CM-2 located at the Supercomputer Computations Research Institute at Florida State University.

We carried out simulations with two flavors of dynamical staggered quarks using the Hybrid Molecular Dynamics algorithm [7]. The lattice size was $16^3 \times 32$ sites and the lattice coupling $\beta = 5.6$. The dynamical quark masses were $am_q = 0.01$ and 0.025. A total of 2000 simulation time units (with the normalization of Ref. [2]) was generated at each value of the quark mass, after thermalization. The $am_q = 0.01$ run started from an equilibrated $16^4$ lattice of our previous runs on the ETA-10, which was doubled in the time direction and then re-equilibrated for 150 trajectories. The $am_q = 0.025$ run was started from the last configuration of the smaller mass run, and then thermalized for 300 trajectories. For $am_q = 0.01$, we recorded lattices for the reconstruction of spectroscopy every 5 HMD time units, for a total of 400 lattices. At $am_q = 0.025$, lattices were stored every 10 time units for a total of 200 lattices.

For our spectrum calculation, we used periodic boundary conditions in the three spatial directions, and antiperiodic boundary conditions in the temporal direction. To calculate hadron propagators, we fixed the gauge in each configuration to the lattice Coulomb gauge using an overrelaxation algorithm[8], and used sources for the quark Green functions which spread out in space uniformly over the spatial simulation volume and were restricted to a single time slice ("wall" sources[9]). Our inversion technique was the conjugate gradient algorithm, using a fast matrix inverter written in CMIS (a low level assembler for the CM-2) by C. Liu[10, 11].

In this work we used two kinds of wall sources. The first of these consisted of a 1 in a selected color component at each site of the source time slice where the $x$, $y$, and $z$ coordinates were all odd. In other words, the source was restricted to a single corner of each $2^4$ hypercube. This is the same source as used in our previous work, and we will refer to it as the "corner" source or C.

In addition to this corner source, we also used a triplet of wall sources. Following Gupta et al. [12] we defined an "even" source which takes the value +1 on every site of the source time slice, and an "odd" source which is +1 on all the even (space odd) sites on the source time slice and −1 on all the odd (space even) sites on this time slice. For definiteness, in this paragraph we take the source time slice to be $t = 1$. These sources allowed calculation of the $\Delta$ propagator and propagators for some of the local and non-local mesons. (We use the $\Delta$ propagator corresponding to a point sink where the three quarks are displaced by one unit in the $x$, $y$, and $z$ directions, respectively, from the origin of the unit cube [13].) The third source we used was what we call a "vector" source. This source is +1 on all sites on the source time slice that have an even $y$ coordinate and −1 for those whose $y$ coordinate is odd. With these three sources we are able to calculate meson propagators for all 20 meson representations of the time slice group which are local in time [14]. In addition, we have calculated the propagator for a local nucleon, and the $\Delta$ discussed above, from the "even" source quark Green functions. We will refer to this triplet of sources as EOV.

For the mesons, we averaged propagators computed from six sets of wall sources at time slices 1, 2, 3, 17, 18 and 19. Propagators from three consecutive time slices were needed for a separate study of glueball to $\bar{q}q$ correlations. For the baryons, we used four wall sources at time slices 1, 9, 17 and 25.

Finally, for comparison, we also measured the hadron propagators from a point source. This calculation was performed "on line" every time unit, for a total of 2000 measurements for each quark mass.

# 3 RESULTS

## 3.1 Doubling effects on the pion propagator

In our previous, work we used $12^4$ and $16^4$ lattices doubled or quadrupled in the time direction for computing propagators. We found irregularities in the effective mass as a function of distance from the source for the pion. (The effective mass is the mass obtained by fitting with zero degrees of freedom to points in the propagator centered at some distance.) For the pion propagator, where we fit to a simple exponential plus the piece from periodicity, the effective mass at distance $T$ is obtained from the two points in the propagator at time distances $T - \frac{1}{2}$ and $T + \frac{1}{2}$. For other particles, where we use four parameter fits, with one particle of each parity, the effective masses are obtained from four successive distances in the propagator ( $T - \frac{3}{2}$, $T - \frac{1}{2}$, $T + \frac{1}{2}$, $T + \frac{3}{2}$). Since the location of these features seemed to depend on the lattice size before doubling, we tentatively ascribed them to effects of the doubling[2]. A simple analytic model of a doubled lattice showed similar features[3]. In the current work, we generated configurations on a $16^3 \times 32$ lattice, and did not double in the time direction when computing the hadron spectrum. The pion propagator is much better behaved. We show the new results for the pion effective mass together with our previous results in Figs. 1 and 2, for $am_q = 0.025$ and 0.01 respectively. Note that one of the sources used in the present work, the "corner" wall source, is identical to the source used in the previous work. We see that the pion effective mass in the current work is the same for the two sources, and is much smoother than on the doubled lattices.

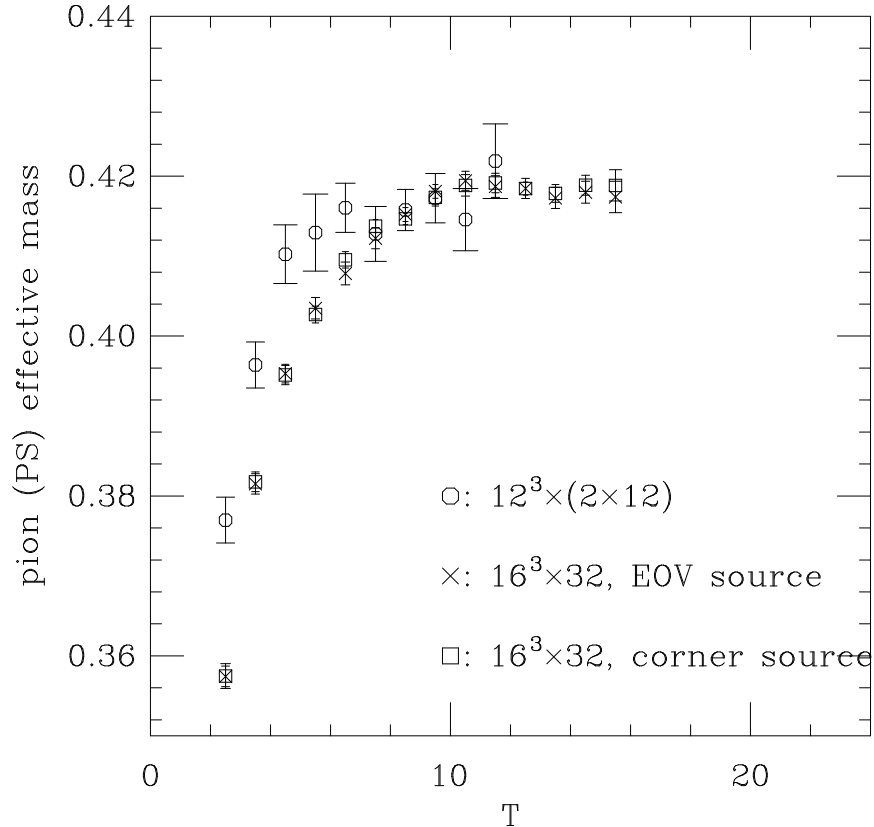

Figure 1: Pion effective mass versus distance for $am_q = 0.025$.

In Figs. 3 and 4, we show the effective mass plots for the $\rho$. Again, we notice that they are relatively flat, in contrast with the work using doubled lattices. [2]

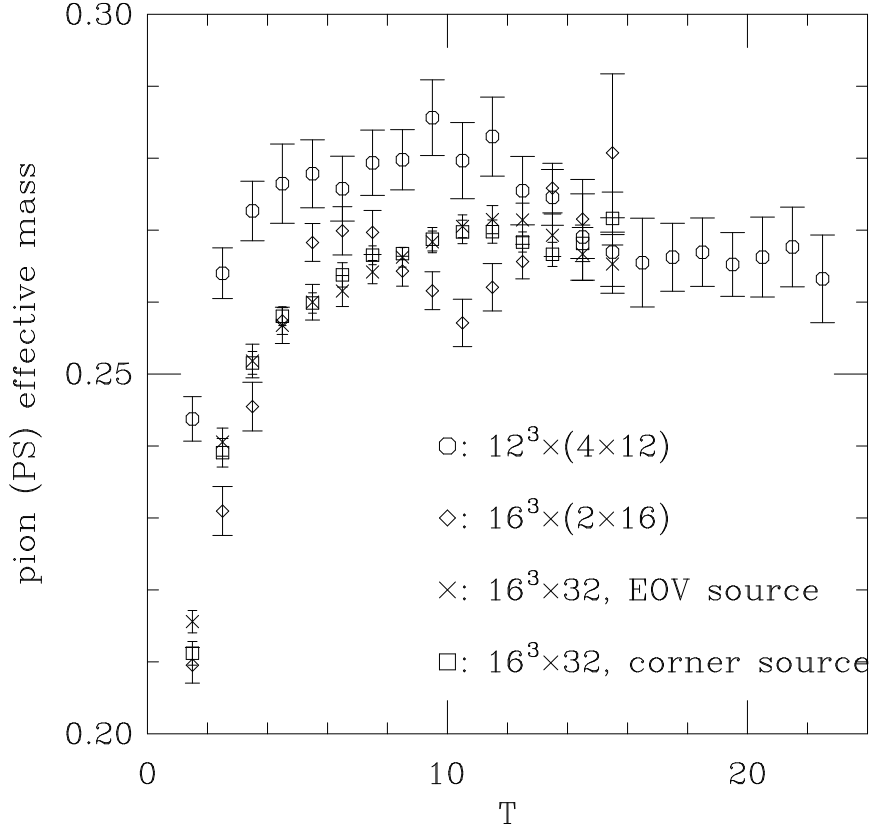

Figure 2: Pion effective mass versus distance for $am_q = 0.01$.

## 3.2  Best estimates for masses

Hadron masses were estimated by making correlated fits to the average propagator[15]. To reduce the effects of autocorrelations in simulation time, propagators on several successive lattices were averaged together before computing the covariance matrix. For example, we most commonly blocked 8 lattices together for $am_q = 0.01$, or 40 time units, since we measured every five time units. For $am_q = 0.025$ we typically blocked together 4 lattices or again 40 time units.

To display the fits we use figures in which the symbol size is proportional to the confidence level of the fits. The symbol size in the keys corresponds to a confidence level of

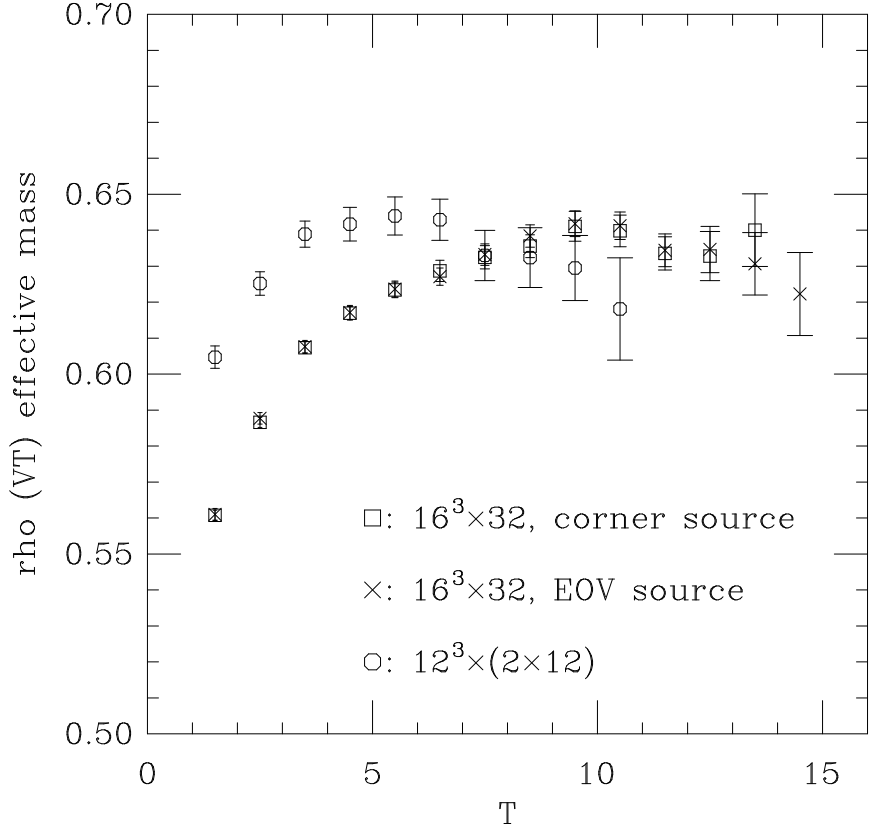

Figure 3: Rho effective mass versus distance for $am_q = 0.025$.

0.5. We plot the fits as a function of the minimum distance used in the fit. To show how the fit quality varies with distance from the source, we plot the fits with two degrees of freedom. Such fits for the $\rho$ masses are displayed in Figs. 5 and 6 and for the nucleon in Figs. 7 and 8.

Tables 1–4 give our estimates for the hadron masses. In the continuum, all 15 components of the $\pi$ multiplet should be degenerate, as should all 15 components of the $\rho$ multiplet. (Although we have only two flavors of quarks in internal lines, the external quark lines have four quark flavors. Hence, in the continuum limit hadrons form multiplets of flavor $SU(4)$.) When using staggered quarks on the lattice, flavor symmetry is explicitly broken, and each continuum flavor multiplet is broken down into irreducible representations of the discrete

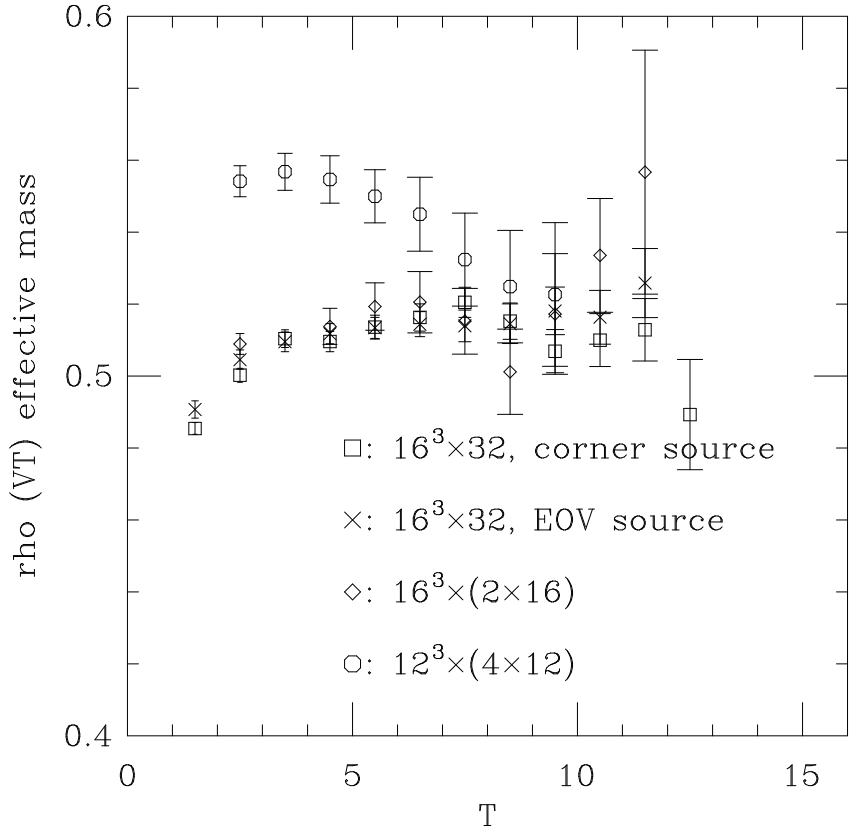

Figure 4: Rho effective mass versus distance for $am_q = 0.01$.

symmetries of the lattice action restricted to a given time slice [14]. Full flavor symmetry should be restored in the continuum limit. The extent to which this symmetry is restored at a finite lattice spacing gives us some indication as to whether our lattice spacing is small enough ($\beta$ large enough) to adequately approximate the continuum limit. In Figs. 9 and 10, we plot the $\pi$ and $\rho$ masses, respectively, from Tables 1–4 for the different representations of the time slice group accessible using the EOV sources. We notice that flavor symmetry appears to be good to a few percent for the $\rho$ multiplet. For the $\pi$ sector there is approximate degeneracy for the non-Goldstone pions, but the mass of the Goldstone pion still lies significantly below that of the rest $((m_{\tilde{\pi}} - m_\pi)/m_\pi \approx 0.3$ for $am_q = 0.01)$. This should not surprise us since, for the quenched approximation, definitive evidence for the restoration of the mass degeneracy between the Goldstone and non-Goldstone pions has only been claimed for $\beta = 6.5$ [16]. In Sec. 3.5, we shall indicate that our $\beta$ (5.6) is more comparable with a quenched system at $\beta \approx 5.95$. The squared Goldstone pion mass is very nearly proportional to $m_q$ (see Sec. 3.4). The other pion masses do not extrapolate to zero with $m_q$. For example, at $am_q = 0.025$ the mass ratio $m_{\tilde{\pi}}/m_\pi = 1.223(5)$, while at $am_q = 0.01$ this ratio is 1.306(11). This contrasts with a four flavor study by the $MT_c$ collaboration[17], in which $m_{\tilde{\pi}}^2$ appears to be proportional to $m_q$.

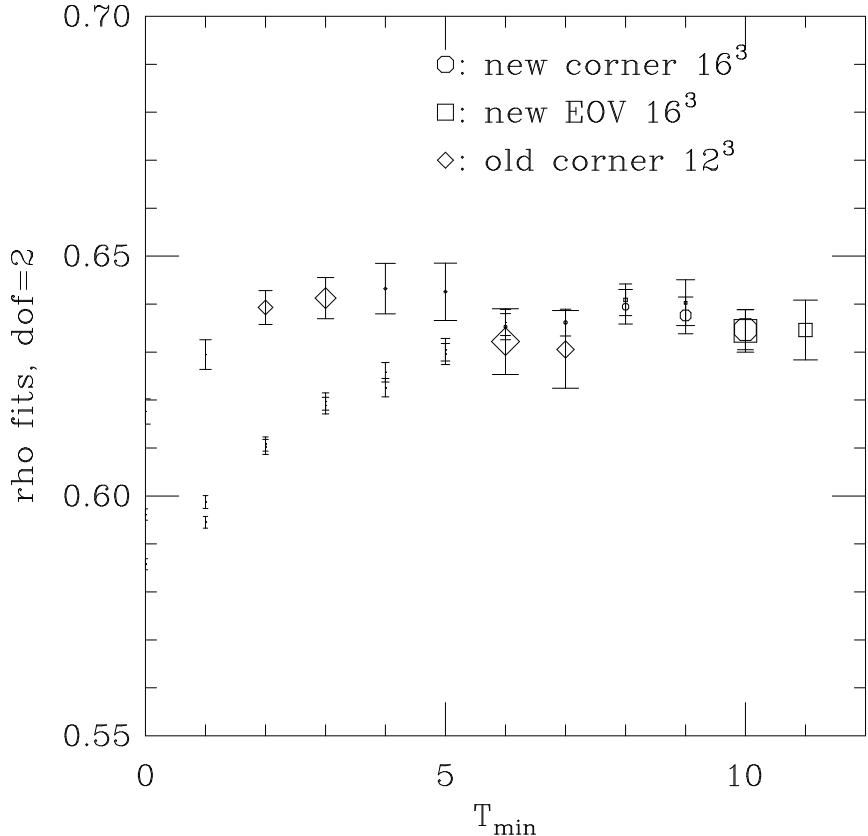

Figure 5: Fits to the $\rho$ mass for $am_q = 0.025$. The size of the points is proportional to the confidence level of the fits.

Fig. 11 gives the "Edinburgh" plot of $m_N/m_\rho$ against $m_\pi/m_\rho$ for the results of Tables 1–4. Fig. 12 is the "Boulder" plot for the $N - \Delta$ mass splitting. Both plots are roughly

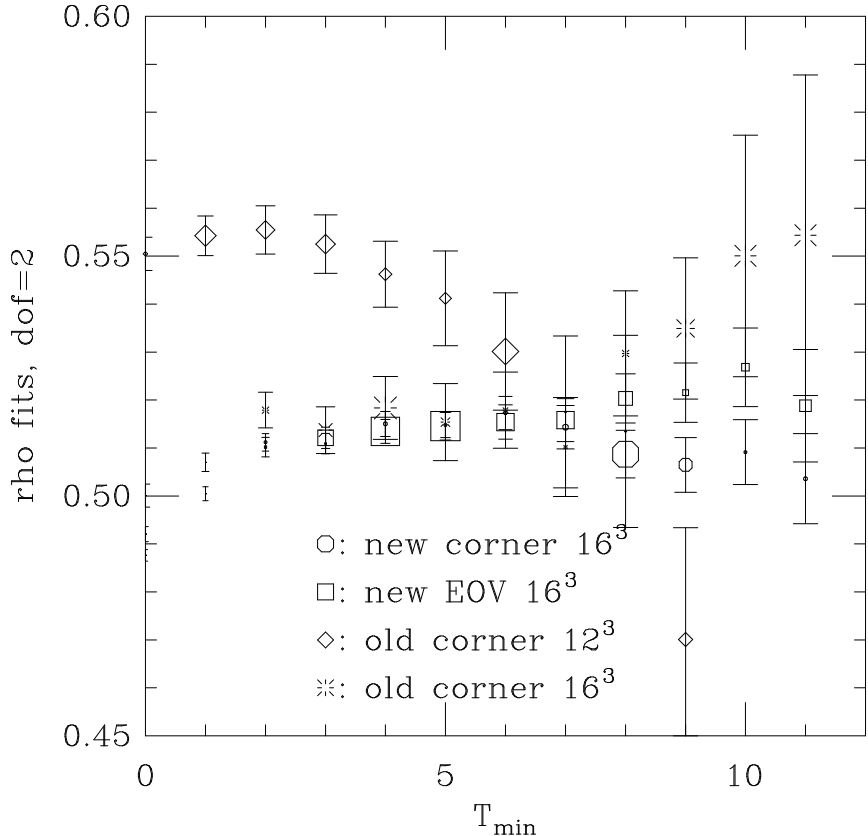

Figure 6: Fits to the $\rho$ mass for $am_q = 0.01$. The size of the points is proportional to the confidence level of the fits.

what one would expect for these values of the quark mass.

## 3.3 Finite size and source effects on the hadron masses

In our earlier work, we found a large change in the nucleon mass with quark mass $am_q = 0.01$ when the spatial lattice size was increased from 12 to 16. With our new results, we can examine this in more detail, as well as extend the study to $am_q = 0.025$.

For $am_q = 0.01$, the nucleon mass on a $12^3 \times (12 \times 2)$ lattice was estimated to be 0.848(11) while that on a $12^3 \times 24$ was found to be 0.815(13), the difference probably being an effect of the doubling. On the $16^3 \times (16 \times 2)$ this had fallen to 0.770(8). In the data of

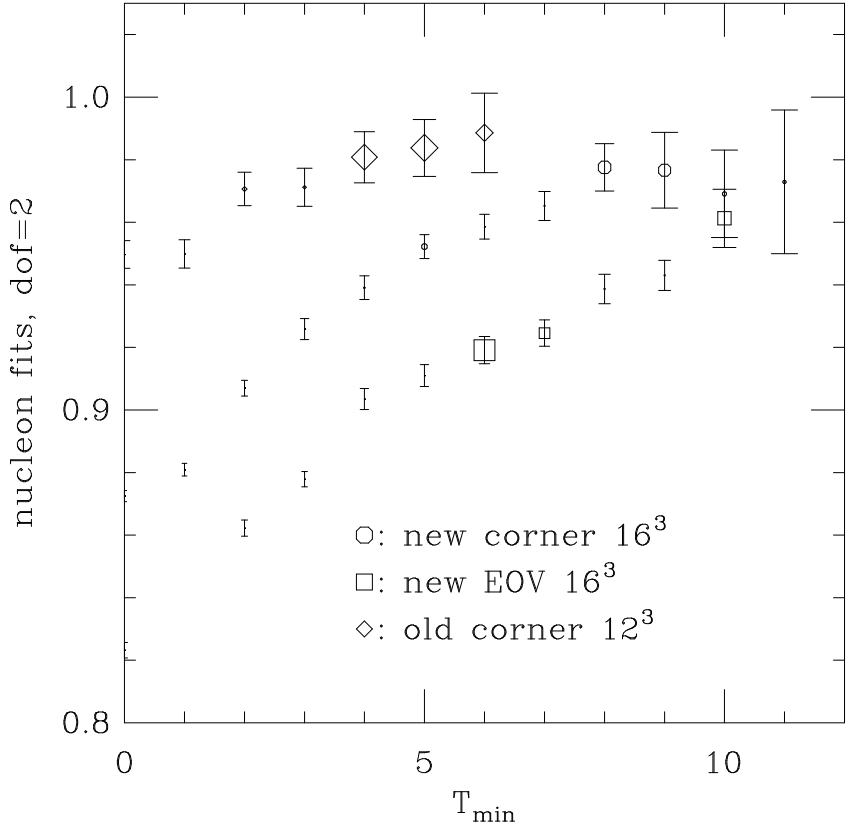

Figure 7: Fits to the nucleon mass for $am_q = 0.025$. The size of the points is proportional to the confidence level of the fits.

Table 1 for the $16^3 \times 32$ lattice we find that for the "corner" source (which is identical to the source used on the smaller lattices) the value is 0.748(4) again lower than the doubled case. Thus, we have further evidence for the finite volume effect reported in [2] and also seen by [18] for the nucleon mass. At $am_q = 0.025$, the nucleon mass on a $12^3 \times (12 \times 2)$ lattice was 0.982(9), while that for a $16^3 \times 32$ lattice (Table 3) is 0.981(8). Thus, it would appear that even a $12^3$ box is adequate to hold a nucleon at $am_q = 0.025$ with no appreciable finite size effects.

For the mesons, we find good agreement between the masses on $12^3 \times 24$, $16^3 \times (16 \times 2)$ and the new results on $16^3 \times 32$ lattices, for both quark masses. The $12^3 \times (12 \times 2)$ $\rho$ effective masses showed no clear plateau, and this is reflected in the other fits, so that the mass values were unreliable. The observed undulating behavior of the $\pi$ effective masses on the $12^3 \times (12 \times 2)$ lattice reflects itself in the more general fit. (The $16^3 \times (16 \times 2)$ lattice shows similar problems.) Within these ambiguities, the new results are in good agreement with those for smaller lattices. Hence we may conclude that there are no significant finite size effects in the meson masses for spatial boxes with volumes $\gtrsim 12^3$ for quark masses $am_q \gtrsim 0.01$ at $\beta = 5.6$.

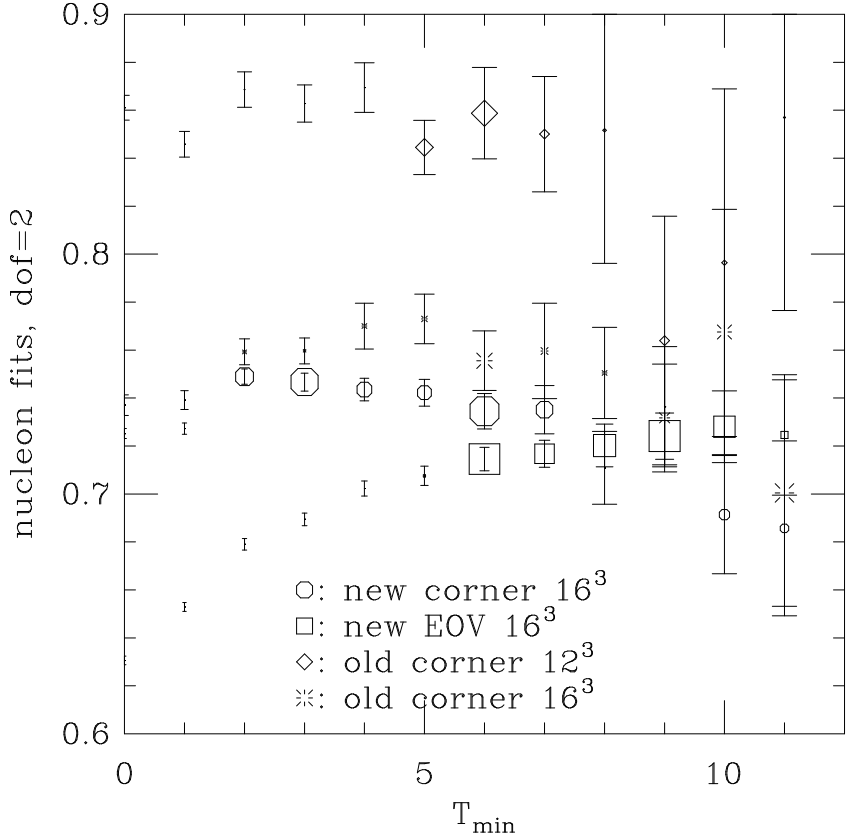

Figure 8: Fits to the nucleon mass for $am_q = 0.01$. The size of the points is proportional to the confidence level of the fits.

Now let us discuss the effects of the two different types of source. For mesons, there is only one wall source, and one point sink corresponding to each component of each irre-ducible representation of the time slice group [14]. This means that for those representations occurring in both the "corner" and EOV wall sources we can expect to get the same results in both cases. This is well born out by the masses of Tables 1–4.

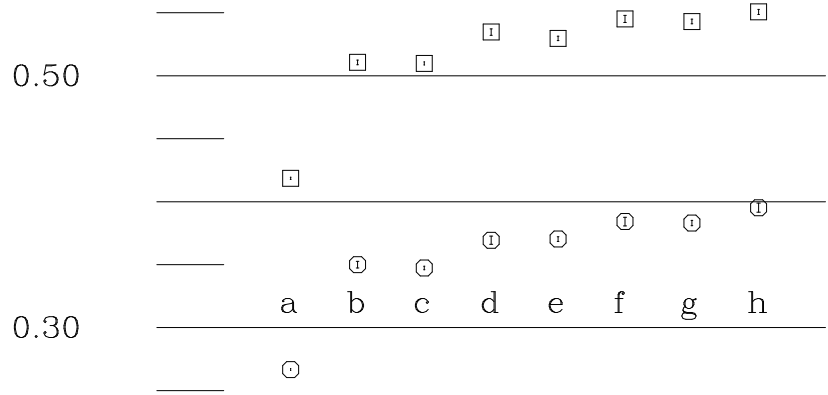

Figure 9: Masses of the various lattice pions. The octagons are for $am_q = 0.01$ and the squares for $am_q = 0.025$. From left to right, the representations are (a) $\pi$ (Goldstone), (b) $\tilde{\pi}$, (c) $\pi^3(1)$, (d) $\tilde{\pi}^3(1)$, (e) $\pi^3(2)$, (f) $\tilde{\pi}^3(2)$, (g) $\pi(3)$ and (h) the $\eta'/\pi$.

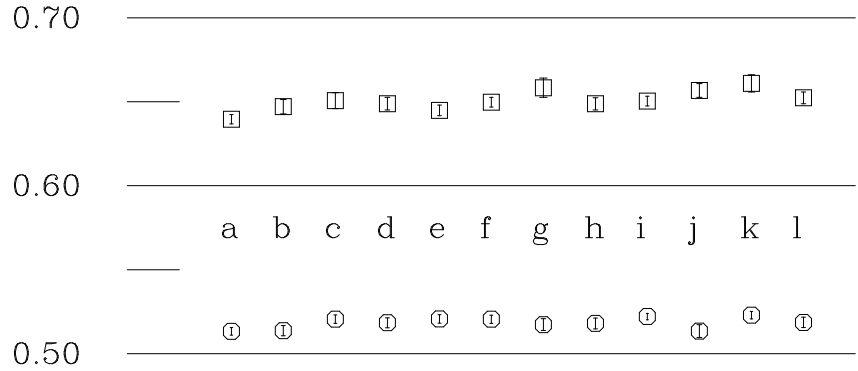

Figure 10: Masses of the various lattice $\rho$ mesons. From left to right the representations are (a) $\rho$ (VT), (b) $\tilde{\rho}$ (PV), (c) $\omega/\rho$, (d) $\tilde{\rho}^3(1)$, (e) $\rho^6(1)$, (f) $\tilde{\rho}^6(1)$, (g) $\rho^3(2)$, (h) $\tilde{\rho}^3(2)$, (i) $\rho^6(2)$ and (j) $\tilde{\rho}^6(2)$. (k) $\rho(3)$ and (l) $\tilde{\rho}(3)$.

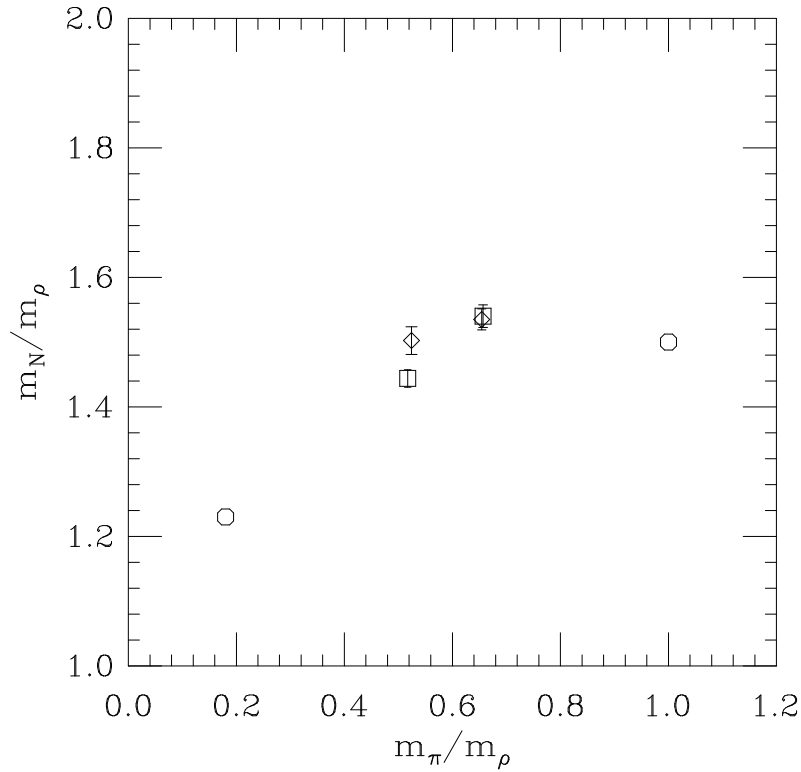

Figure 11: Edinburgh plot. The diamonds are our results from the previous simulations, and the squares are the new results. Error bars are the statistical errors only, and do not include uncertainty based on the choice of source or fitting range. (We use the even-odd source results here.)

For the nucleon, we use a local sink which projects the **8** representation of the time slice group. The "corner" source produces only one baryon representation, the local **8** representation. The "even" source, on the other hand produces all baryon representations, and in particular 5 copies of the **8** representation. Only one of these **8**'s is local; the other 4 have quarks on more than 1 vertex of the unit cube. The local point sink has overlap

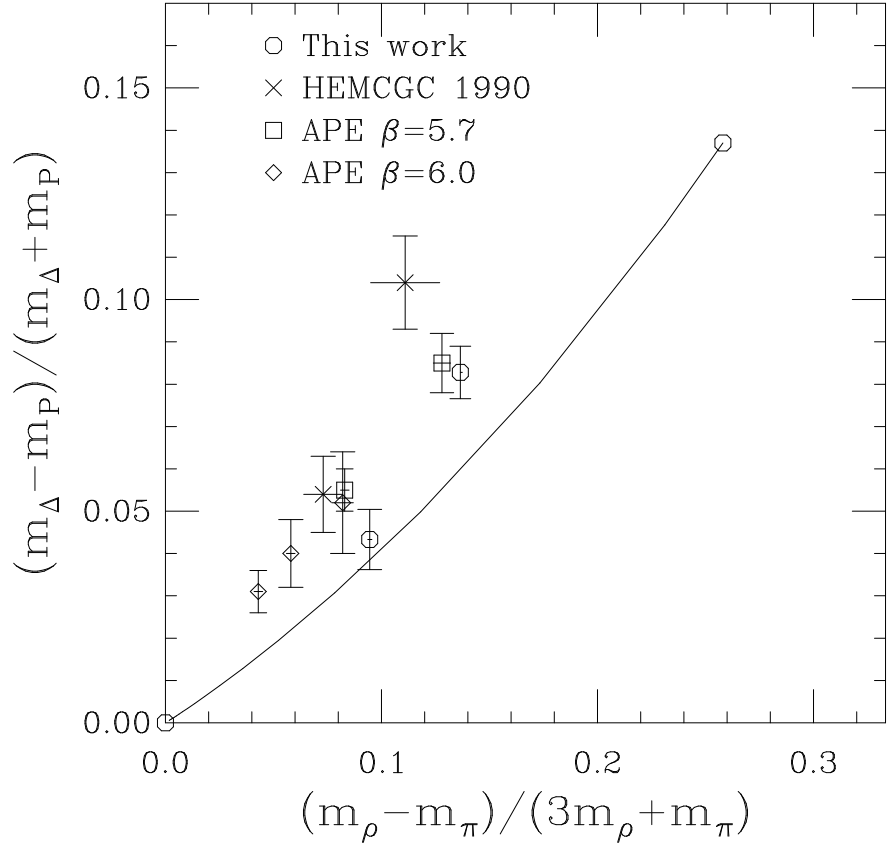

Figure 12: Comparison of baryon and meson hyperfine splitting. The two circles show the expected values of hyperfine splitting in the limit of infinite quark mass and from experiment; the line interpolating between them is a simple quark model. The APE data is from a quenched simulation[19].

with to all 5 of these octets, each of which will, in general, have different couplings to the allowed baryon states. For this reason the nucleon propagator for the "even" source can be rather different from that for the corner source. That this is so is illustrated by looking at the effective mass plots (Figs. 13 and 14) for the 2 different nucleon propagators. At $am_q = 0.025$, the effective masses for the "corner" source lie consistently higher than those for the "even" source. Since it is difficult to find strong evidence for a plateau in this data (at least not for the "corner source") the problem could well be that the plateau starts just as the signal/noise ratio starts to worsen. In any case, our best fits (Table 3) are within

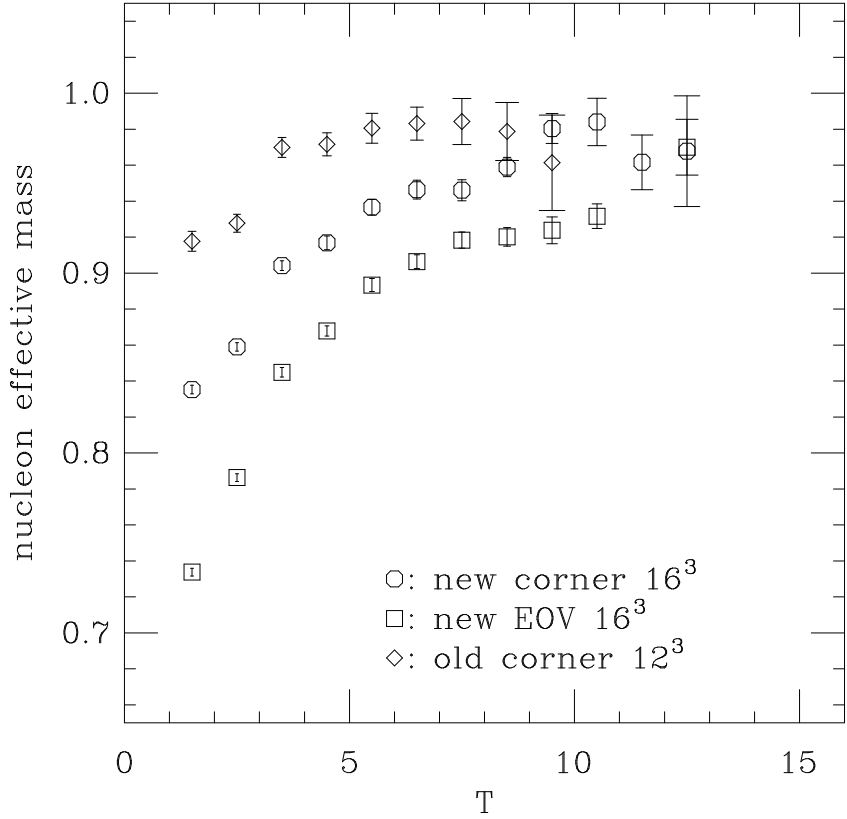

Figure 13: Nucleon effective mass versus distance for $am_q = 0.025$.

2 standard deviations of one another, and can thus be considered to be consistent. For $am_q = 0.01$, the effective masses for the "corner" nucleon again lie consistently above those for the "even" source. However, in the graph of Fig. 14, one notes that the effective masses for the 2 sources appear to be coming together for $T \gtrsim 7.5$. If this is correct, the reason for the discrepancy between the two estimates of the nucleon mass is that our fitting criterion favors the false plateau $3.5 \lesssim T \lesssim 6.5$ in the nucleon effective mass plot. If we had better statistics we would presumably find the true plateau.

Finally let us comment on the point source fits as compared with the wall fits. Only in the case of the $\pi$ do these point source fits have the quality of the wall fits. The $\pi$ masses obtained from the point and wall sources are in excellent agreement. For the other particles, the rapid decrease of the point source propagators with increasing $T$ due to contamination with higher mass excitations produces mass estimates that tend to be high and at the very least have much larger errors than the wall results. The rapid decrease of effective masses with $T$ for the point sources makes the evidence that these reach a plateau before the signal is lost less compelling than in the case of wall sources. Their main virtue is that their mass predictions give an upper bound on the particle mass. This is no great advantage if the bound is too large or has a very large uncertainty (error).

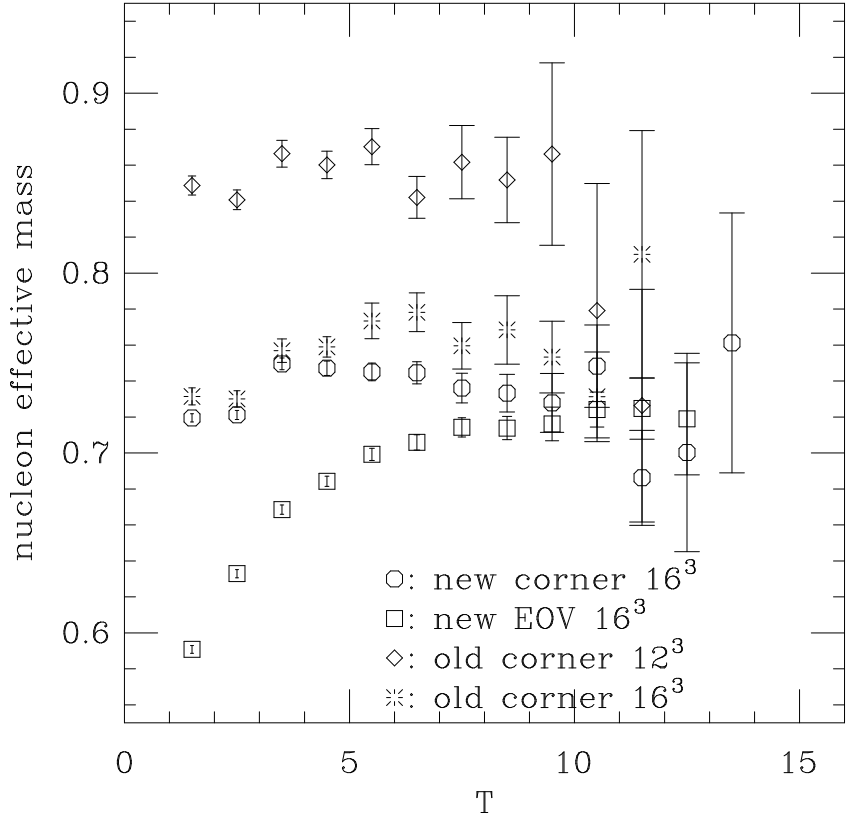

Figure 14: Nucleon effective mass versus distance for $am_q = 0.01$.

Eventually, lattice QCD should provide masses for excited state hadrons as well as for the lowest mass particle for given quantum numbers. Identification of excited states is probably easiest in the pion channnel, since the small ground state mass means that the excited state is probably well separated in mass from the ground state. In Fig. 15, we show fits to the pion from Euclidean time range $T_{min}$ to 16, including both the ground state and excited state masses for the two particle fits (both with the same parity). In these graphs the symbol size is proportional to the confidence level of the fits. At $am_q = 0.025$ we see consistent results for the excited state mass, independent of $T_{min}$ up to the point where the excited state is no longer needed in the fit, and independent of whether we use the EOV or the corner wall source. For the point source, we see that including two particles is still not sufficient to get good fits with small $T_{min}$. Unfortunately, for $am_q = 0.01$ the results are not nearly as good, although this is one of the cases for which we have no good fits even with large $T_{min}$.

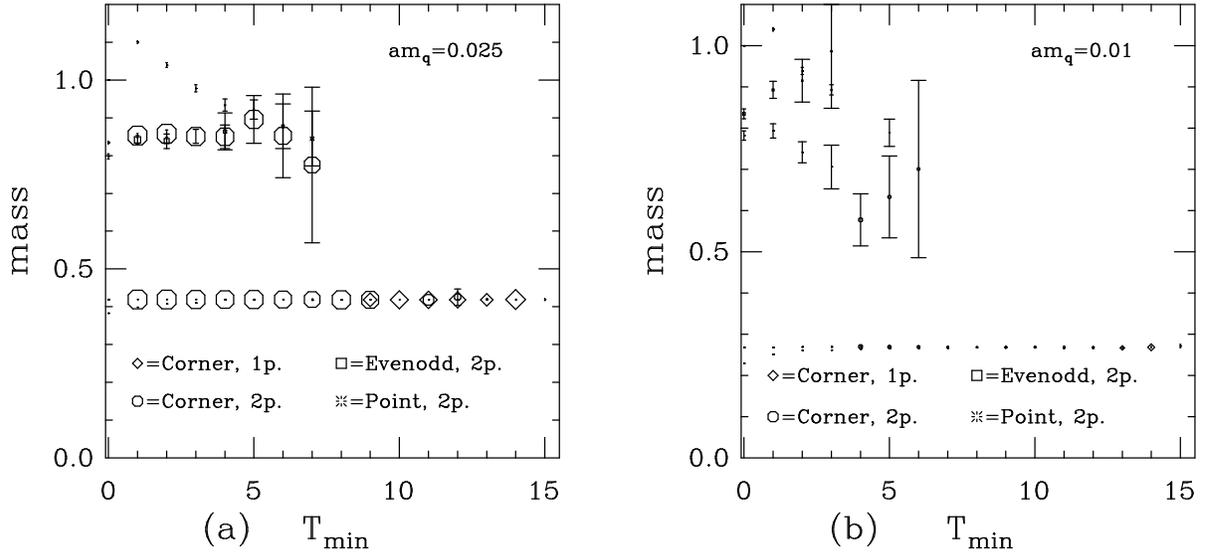

Figure 15: Goldstone pion fits including excited state masses. The symbol size is proportional to the confidence level, with the symbol size used in the legend corresponding to 50% confidence. We show results for $am_q = 0.025$ (a) and 0.01 (b). The octagons, squares and bursts correspond to two particle fits, both particles having the same parity, with corner, EOV and point sources respectively, while the diamonds are from one particle fits to the corner wall source.

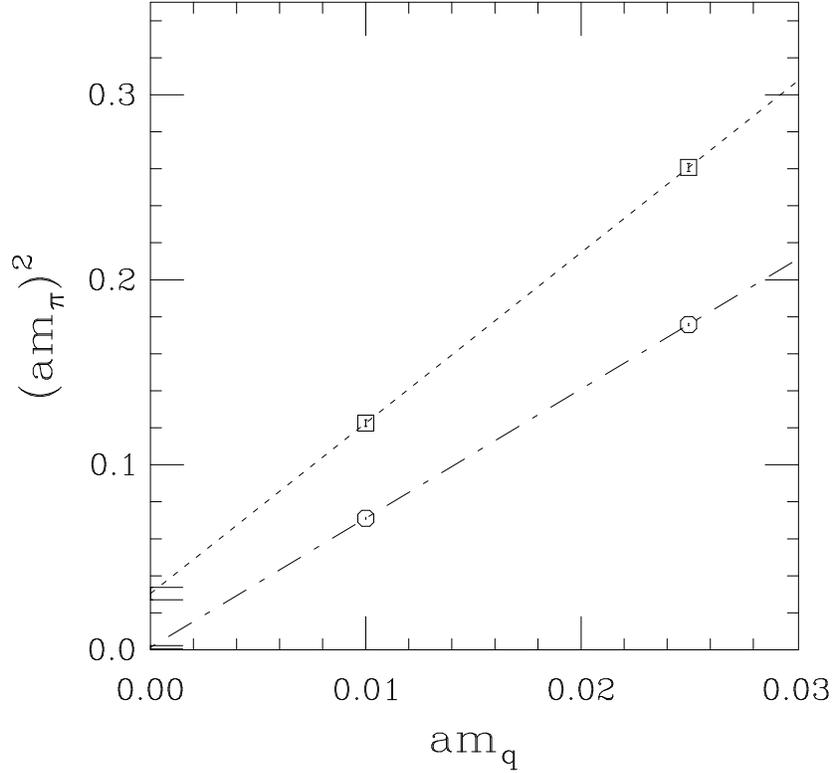

Figure 16: Squared pion masses versus quark mass. The octagons are for the Goldstone pion, and the squares for the other pointlike pion, the $\tilde{\pi}$. The dashed and dotted lines are extrapolations to zero quark mass, and the horizontal lines on the left side are one standard deviation limits on this extrapolation. (In reality, we do not expect the $\tilde{\pi}$ mass squared to be vanish linearly with $m_q$ for small $m_q$.)

## 3.4 PCAC

PCAC predicts that $m_\pi^2 \propto m_q$. In Fig. 16, we plot $m_\pi^2$ against $m_q$. For definiteness we have chosen the 4 parameter EOV estimate for the pion mass in each case. We obtain

$$m_\pi^2 = 0.0013(9) + 6.98(5) m_q.$$

The intercept is only $1.4\sigma$ from zero. Thus, this simple PCAC relationship appears to be well satisfied. We therefore can make use of the the more precise relationship

$$f_\pi^2 m_\pi^2 = m_q \langle \bar{\psi}\psi \rangle$$

for $m_q$ sufficiently small, to extract an estimate for $f_\pi$. To finess the question of perturbative subtractions for $\langle \bar{\psi}\psi \rangle$ which are known to remove most of its apparent mass dependence, we linearly extrapolate it to $m_q = 0$ where no such subtraction is necessary. Our measured values for $a^3 \langle \bar{\psi}\psi \rangle$ were $0.11223(46)$ at $am_q = 0.01$ and $0.21398(34)$ for $am_q = 0.025$. This gives $a^3 \langle \bar{\psi}\psi \rangle = 0.04440(80)$ at $m_q = 0$. Since $\langle \bar{\psi}\psi \rangle$ is measured with 4 fermion flavors, we must multiply it by $(N_f = 2)/4 = 0.5$ before inserting it into the above equation giving

$$af_\pi = 0.0564(5).$$

Estimating $a$ by setting the $\rho$ mass linearly extrapolated to $m_q = 0$ to its experimental value we find $a = 1/(1.80(2)\text{GeV})$ (from the EOV $\rho$). This gives us

$$f_\pi = 102(2) MeV$$

as compared with the experimental value $f_\pi \approx 93 MeV$. Note that the error we have quoted represents only the statistical error. Just taking into account the systematic uncertainty of choosing which $\rho$ masses to use, indicates that the error estimate should almost certainly be at least twice what we have quoted. With this in mind, and remembering that we really have little justification for linearly extrapolating our $\rho$ masses to $m_q = 0$, we consider this value to be quite good.

## 3.5 Comparison between quenched and full QCD

In addition to simulating with 2 flavors of dynamical quarks, we also estimated the hadron spectrum for quenched QCD on a $16^3 \times 32$ lattice. Here the aim was to compare the spectrum from quenched QCD with that of full QCD. For this reason, we chose $\beta$ values for the quenched runs which we believed to be in close correspondence with $\beta = 5.6$ for full (2 flavor) QCD. Two values of $\beta$ were chosen so that interpolation to a requisite $\beta$ might be possible. We chose $\beta = 5.85$ and $\beta = 5.95$ as our two values. The hadron masses for these $\beta$'s at $am_q = 0.01$ and $am_q = 0.025$ are given in Tables 5–6. A cursory comparison of the masses in these tables with those for the full theory (Tables 1–4) shows why these two $\beta$ values were chosen. At $\beta = 5.85$ the masses (in lattice units) of the Goldstone pions are very close to their values in the full theory for both quark masses. At $\beta = 5.95$, on the other hand, the masses of the $\rho$ and nucleon at $am_q = 0.01$ are close to the values in full QCD.

What is immediately clear from these results is that a simple shift in the coupling constant (*i.e.*, in $\beta$) is inadequate to reproduce the whole effect of including dynamical quarks as some have suggested. However, a $\beta$ shift combined with a renormalization of the bare quark mass can bring the spectra into reasonable agreement. We find that the quenched spectrum at $\beta = 5.95$ is brought into reasonable agreement with the dynamical quark spectrum at $\beta = 5.6$ if we increase the bare masses in the quenched case by a factor of 1.16. The reason we must scale all quark masses by the same factor is because both the quenched and the dynamical Goldstone pions appear to obey PCAC. The new Goldstone pion masses are obtained using PCAC from those at $am_q = 0.01$ and $am_q = 0.025$. The non-Goldstone pion mass is obtained by noting that the difference between the non-Goldstone and Goldstone pion masses depends only weakly on the mass. Rho and nucleon masses for the "renormalized" masses are obtained from those at $am_q = 0.01$ and $am_q = 0.025$ by linear interpolation/extrapolation. The comparison between these quenched and unquenched masses is exhibited in Tables 7–8. These results might have been improved still further if we had varied $\beta$ in the neighborhood of $\beta = 5.95$. The mass differences between the full QCD and quenched masses in Tables 7–8 are larger than can be attributed to statistics alone, but are probably consistent with the systematic errors due to choices of fits and differences in the finite size/lattice spacing errors between the two theories. It therefore remains to be seen whether there are significant differences in the infinite volume continuum theories.

# ACKNOWLEDGEMENTS

This work was supported by the U. S. Department of Energy under contracts DE–FG05–92ER–40742, DE–FG02–85ER–40213, DE–AC02–86ER–40253, DE–AC02–84ER–40125, FG–02–91ER–40661, W-31-109-ENG-38, and by the National Science Foundation under grants NSF-PHY87-01775, and NSF-PHY91-16964. The computations were carried out at the Florida State University Supercomputer Computations Research Institute which is partially funded by the U.S. Department of Energy through Contract No. DE-FC05-85ER250000. One of us (DKS) would like to thank S. Kim for familiarizing him with the usage of the CERNLIB minimization routine MINUIT, which was used in obtaining some of the correlated fits, and also to thank him and G. T. Bodwin for many helpful discussions. We thank Jim Hudgens, for his assistance, and T. Kitchens and J. Mandula for their continuing support and encouragement.

| $am_q = 0.01$ | | | | | | | |
|---|---|---|---|---|---|---|---|
| particle | source | range | mass | error | $\chi^2/d.o.f.$ | confidence | parameters |
| $\pi$ | point | 12–16 | 0.2681 | 0.0010 | 0.13 | 0.95 | 2 |
| $\pi$ | EOV | 7–16 | 0.2667 | 0.0008 | 1.90 | 0.55 | 2 |
| $\pi$ | EOV | 1–16 | 0.2673 | 0.0008 | 1.50 | 0.12 | 4 |
| $\pi$ | C | 13–16 | 0.2667 | 0.0015 | 1.45 | 0.24 | 2 |
| $\pi$ | C | 4–16 | 0.2700 | 0.0012 | 1.33 | 0.20 | 4 |
| $\tilde{\pi}$ | point | 9–16 | 0.3899 | 0.0190 | 1.27 | 0.28 | 4 |
| $\tilde{\pi}$ | EOV | 6–16 | 0.3500 | 0.0026 | 0.885 | 0.52 | 4 |
| $\tilde{\pi}$ | C | 9–16 | 0.3553 | 0.0039 | 0.20 | 0.94 | 4 |
| $\pi^3(1)$ | EOV | 5–16 | 0.3474 | 0.0014 | 1.99 | 0.30 | 2 |
| $\tilde{\pi}^3(1)$ | EOV | 6–16 | 0.3694 | 0.0032 | 0.77 | 0.61 | 4 |
| $\pi^3(2)$ | EOV | 7–16 | 0.3703 | 0.0019 | 1.25 | 0.19 | 2 |
| $\tilde{\pi}^3(2)$ | EOV | 8–16 | 0.3842 | 0.0034 | 1.30 | 0.26 | 4 |
| $\pi(3)$ | EOV | 7–16 | 0.3831 | 0.0021 | 0.89 | 0.52 | 2 |
| $\eta'/\pi$ | EOV | 6–16 | 0.3952 | 0.0036 | 0.44 | 0.88 | 4 |
| $\rho$ | point | 9–16 | 0.492 | 0.038 | 0.65 | 0.63 | 4 |
| $\rho$ | EOV | 3–16 | 0.5133 | 0.0022 | 1.11 | 0.35 | 4 |
| $\rho$ | C | 8–16 | 0.5085 | 0.0050 | 1.54 | 0.17 | 4 |
| $\tilde{\rho}$ | point | 8–16 | 0.476 | 0.035 | 0.31 | 0.91 | 4 |
| $\tilde{\rho}$ | EOV | 4–16 | 0.5152 | 0.0032 | 1.74 | 0.07 | 4 |
| $\tilde{\rho}$ | C | 10–16 | 0.4918 | 0.0091 | 0.40 | 0.74 | 4 |
| $\omega/\rho$ | EOV | 2–16 | 0.5206 | 0.0025 | 1.31 | 0.21 | 4 |
| $\tilde{\rho}^3(1)$ | EOV | 2–16 | 0.5184 | 0.0030 | 0.85 | 0.59 | 4 |
| $\rho^6(1)$ | EOV | 4–16 | 0.5207 | 0.0025 | 0.64 | 0.76 | 4 |
| $\tilde{\rho}^6(1)$ | EOV | 3–16 | 0.5205 | 0.0024 | 1.13 | 0.34 | 4 |
| $\rho^3(2)$ | EOV | 5–16 | 0.5173 | 0.0035 | 1.05 | 0.39 | 4 |
| $\tilde{\rho}^3(2)$ | EOV | 2–16 | 0.5180 | 0.0032 | 0.92 | 0.52 | 4 |
| $\rho^6(2)$ | EOV | 3–16 | 0.5221 | 0.0019 | 0.88 | 0.55 | 4 |
| $\tilde{\rho}^6(2)$ | EOV | 5–16 | 0.5134 | 0.0038 | 1.29 | 0.24 | 4 |
| $\rho(3)$ | EOV | 2–16 | 0.5229 | 0.0022 | 0.75 | 0.69 | 4 |
| $\tilde{\rho}(3)$ | EOV | 3–16 | 0.5186 | 0.0031 | 0.73 | 0.70 | 4 |
| $N$ | point | 9–14 | 0.738 | 0.086 | 0.47 | 0.63 | 4 |
| $N$ | EOV | 7–15 | 0.720 | 0.006 | 0.36 | 0.88 | 4 |
| $N$ | C | 10–15 | 0.696 | 0.027 | 2.00 | 0.37 | 4 |
| $N$ | C | 0–15 | 0.727 | 0.008 | 0.31 | 0.96 | 8 |
| $N'$ | point | NP | NP | NP | NP | NP | 4 |
| $N'$ | point | 7–13 | 1.209 | 0.087 | 1.21 | 0.30 | 4 |
| $N'$ | EOV | 6–15 | 0.948 | 0.066 | 0.37 | 0.90 | 4 |
| $N'$ | EOV | 0–15 | 0.948 | 0.025 | 0.31 | 0.96 | 8 |
| $N'$ | C | 3–15 | 0.904 | 0.009 | 1.78 | 0.06 | 4 |
| $\Delta$ | EOV | 4–15 | 0.850 | 0.008 | 0.39 | 0.93 | 4 |
| $\Delta'$ | EOV | 6–15 | 1.031 | 0.065 | 0.47 | 0.83 | 4 |

Table 1: Hadron masses $am_q = 0.01$. Notation: superscript is dimension of representation of time slice group; number of links in parenthesis; tilde (˜) state has extra $\gamma_0$; notation abbreviated when unambiguous. "NP" indicates no plateau was found in the mass fits.

| $am_q = 0.01$ | | | | | | | |
|---|---|---|---|---|---|---|---|
| particle | source | range | mass | error | $\chi^2/d.o.f.$ | confidence | parameters |
| $\pi^*$ | EOV | 1–16 | 0.893 | 0.021 | 1.50 | 0.12 | 4 |
| $\pi^*$ | C | 4–16 | 0.578 | 0.063 | 1.33 | 0.20 | 4 |
| $\pi^*$ | point | 5–16 | 0.789 | 0.033 | 1.87 | 0.06 | 4 |
| $f_0/a_0$ | point | 10–16 | 0.547 | 0.015 | 1.61 | 0.19 | 4 |
| $f_0/a_0$ | EOV | 6–16 | 0.514 | 0.008 | 0.89 | 0.52 | 4 |
| $f_0/a_0$ | C | 7–16 | 0.505 | 0.013 | 0.78 | 0.59 | 4 |
| $a_0^3(1)$ | EOV | 6–16 | 0.615 | 0.014 | 0.77 | 0.61 | 4 |
| $a_0^3(2)$ | EOV | 6–16 | 0.615 | 0.019 | 1.53 | 0.15 | 4 |
| $a_0(3)$ | EOV | 6–16 | 0.645 | 0.020 | 0.44 | 0.88 | 4 |
| $a_1$ | point | 8–16 | 0.683 | 0.097 | 0.31 | 0.91 | 4 |
| $a_1$ | EOV | 5–16 | 0.700 | 0.011 | 1.88 | 0.06 | 4 |
| $a_1$ | C | 6–16 | 0.744(?) | 0.020 | 2.71 | 0.008 | 4 |
| $a_1^3(1)$ | EOV | 3–16 | 0.693 | 0.007 | 0.93 | 0.51 | 4 |
| $a_1^6(1)$ | EOV | 6–16 | 0.712 | 0.013 | 1.18 | 0.31 | 4 |
| $a_1^3(2)$ | EOV | 5–16 | 0.655 | 0.018 | 0.57 | 0.80 | 4 |
| $a_1^6(2)$ | EOV | 3–16 | 0.701 | 0.004 | 1.27 | 0.24 | 4 |
| $a_1(3)$ | EOV | 4–16 | 0.726 | 0.011 | 0.80 | 0.62 | 4 |
| $b_1$ | point | 7–16 | 0.818 | 0.135 | 0.95 | 0.46 | 4 |
| $b_1$ | EOV | 3–16 | 0.686 | 0.008 | 1.11 | 0.35 | 4 |
| $b_1$ | C | 6–16 | 0.775(?) | 0.042 | 1.86 | 0.08 | 4 |
| $b_1^3(1)$ | EOV | 2–16 | 0.719 | 0.007 | 1.31 | 0.21 | 4 |
| $b_1^6(1)$ | EOV | 5–16 | 0.739 | 0.015 | 0.63 | 0.76 | 4 |
| $h_1/b_1$ | EOV | 5–16 | 0.732 | 0.019 | 1.05 | 0.39 | 4 |
| $b_1^6(2)$ | EOV | 2–16 | 0.717 | 0.004 | 1.05 | 0.40 | 4 |
| $b_1(3)$ | EOV | 2–16 | 0.711 | 0.006 | 0.75 | 0.69 | 4 |

Table 2: Hadron masses $am_q = 0.01$. Notation: superscript is dimension of representation of time slice group; number of links in parenthesis; tilde ( ̃) state has extra $\gamma_0$; notation abbreviated when unambiguous. The "?" denotes cases where none of the fits were good. The $\pi^*$ is an excited state in the pion channel.

| particle | source | range | mass | error | $\chi^2/d.o.f.$ | confidence | parameters |
|---|---|---|---|---|---|---|---|
| $am_q = 0.025$ | | | | | | | |
| $\pi$ | point | 13–16 | 0.4188 | 0.0005 | 1.67 | 0.17 | 2 |
| $\pi$ | point | 4–16 | 0.4190 | 0.0005 | 1.68 | 0.09 | 4 |
| $\pi$ | EOV | 10–16 | 0.4185 | 0.0009 | 0.90 | 0.48 | 2 |
| $\pi$ | EOV | 1–16 | 0.4193 | 0.0007 | 1.13 | 0.33 | 4 |
| $\pi$ | C | 9–16 | 0.4185 | 0.0006 | 0.63 | 0.70 | 2 |
| $\pi$ | C | 1–16 | 0.4192 | 0.0006 | 0.45 | 0.94 | 4 |
| $\tilde{\pi}$ | point | 9–16 | 0.5120 | 0.0061 | 1.43 | 0.22 | 4 |
| $\tilde{\pi}$ | EOV | 8–16 | 0.5106 | 0.0018 | 1.05 | 0.39 | 4 |
| $\tilde{\pi}$ | C | 10–16 | 0.5089 | 0.0027 | 0.63 | 0.59 | 4 |
| $\pi^3(1)$ | EOV | 9–16 | 0.5098 | 0.0012 | 0.72 | 0.64 | 2 |
| $\tilde{\pi}^3(1)$ | EOV | 8–16 | 0.5347 | 0.0025 | 1.20 | 0.30 | 4 |
| $\pi^3(2)$ | EOV | 9–16 | 0.5297 | 0.0017 | 0.42 | 0.87 | 2 |
| $\tilde{\pi}^3(2)$ | EOV | 9–16 | 0.5451 | 0.0028 | 0.96 | 0.43 | 4 |
| $\pi(3)$ | EOV | 9–16 | 0.5431 | 0.0018 | 0.69 | 0.66 | 2 |
| $\eta'/\pi$ | EOV | 6–16 | 0.5507 | 0.0020 | 0.53 | 0.82 | 4 |
| $\rho$ | point | 10–16 | 0.6243 | 0.0127 | 0.40 | 0.75 | 4 |
| $\rho$ | EOV | 7–16 | 0.6396 | 0.0028 | 1.33 | 0.24 | 4 |
| $\rho$ | C | 8–16 | 0.6396 | 0.0037 | 1.00 | 0.42 | 4 |
| $\tilde{\rho}$ | point | NP | NP | NP | NP | NP | 4 |
| $\tilde{\rho}$ | EOV | 8–16 | 0.6471 | 0.0043 | 0.77 | 0.57 | 4 |
| $\tilde{\rho}$ | C | 6–16 | 0.6437 | 0.0037 | 0.60 | 0.76 | 4 |
| $\omega/\rho$ | EOV | 8–16 | 0.6507 | 0.0047 | 0.45 | 0.81 | 4 |
| $\tilde{\rho}^3(1)$ | EOV | 6–16 | 0.6489 | 0.0038 | 1.98 | 0.05 | 4 |
| $\rho^6(1)$ | EOV | 8–16 | 0.6449 | 0.0031 | 0.92 | 0.47 | 4 |
| $\tilde{\rho}^6(1)$ | EOV | 7–16 | 0.6498 | 0.0029 | 0.64 | 0.70 | 4 |
| $\rho^3(2)$ | EOV | 9–16 | 0.6584 | 0.0057 | 1.28 | 0.28 | 4 |
| $\tilde{\rho}^3(2)$ | EOV | 6–16 | 0.6488 | 0.0036 | 0.58 | 0.77 | 4 |
| $\rho^6(2)$ | EOV | 8–16 | 0.6505 | 0.0029 | 0.48 | 0.79 | 4 |
| $\tilde{\rho}^6(2)$ | EOV | 8–16 | 0.6568 | 0.0041 | 0.77 | 0.57 | 4 |
| $\rho(3)$ | EOV | 9–16 | 0.6610 | 0.0051 | 1.24 | 0.29 | 4 |
| $\tilde{\rho}(3)$ | EOV | 7–16 | 0.6524 | 0.0035 | 2.95 | 0.07 | 4 |
| $N$ | point | 10–15 | 0.926 | 0.028 | 0.17 | 0.84 | 4 |
| $N$ | EOV | 2–15 | 0.949 | 0.010 | 1.83 | 0.10 | 8 |
| $N$ | C | 8–15 | 0.979 | 0.008 | 1.40 | 0.23 | 4 |
| $N'$ | point | NP | NP | NP | NP | NP | 4 |
| $N'$ | EOV | 2–15 | 1.289 | 0.078 | 1.83 | 0.10 | 8 |
| $N'$ | C | 8–15 | 1.137 | 0.089 | 1.40 | 0.23 | 4 |
| $\Delta$ | EOV | 6–15 | 1.035 | 0.010 | 0.33 | 0.92 | 4 |
| $\Delta'$ | EOV | 6–15 | 1.302 | 0.070 | 0.33 | 0.92 | 4 |

Table 3: Hadron masses $am_q = 0.025$. Notation: superscript is dimension of representation of time slice group; number of links in parenthesis; tilde (˜) state has extra $\gamma_0$; notation abbreviated when unambiguous.

| $am_q = 0.025$ | | | | | | | |
|---|---|---|---|---|---|---|---|
| particle | source | range | mass | error | $\chi^2/d.o.f.$ | confidence | parameters |
| $\pi^*$ | point | 7–16 | 0.845 | 0.072 | 1.41 | 0.21 | 4 |
| $\pi^*$ | EOV | 1–16 | 0.842 | 0.012 | 1.13 | 0.33 | 4 |
| $\pi^*$ | C | 1–16 | 0.853 | 0.006 | 0.45 | 0.94 | 4 |
| $f_0/a_0$ | point | 9–16 | 0.696 | 0.007 | 1.43 | 0.22 | 4 |
| $f_0/a_0$ | EOV | 8–16 | 0.699 | 0.010 | 1.05 | 0.39 | 4 |
| $f_0/a_0$ | C | 8–16 | 0.697 | 0.015 | 1.40 | 0.23 | 4 |
| $a_0(1)$ | EOV | 8–16 | 0.848 | 0.030 | 1.20 | 0.30 | 4 |
| $a_0(2)$ | EOV | 6–16 | 0.827 | 0.018 | 1.36 | 0.22 | 4 |
| $a_0(3)$ | EOV | 6–16 | 0.829 | 0.020 | 0.53 | 0.82 | 4 |
| $a_1$ | point | NP | NP | NP | NP | NP | 4 |
| $a_1$ | EOV | 6–16 | 0.886 | 0.022 | 1.38 | 0.21 | 4 |
| $a_1$ | C | 6–16 | 0.892 | 0.021 | 0.60 | 0.76 | 4 |
| $a_1^3(1)$ | EOV | 7–16 | 0.887 | 0.036 | 1.74 | 0.11 | 4 |
| $a_1^6(1)$ | EOV | 6–16 | 0.905 | 0.018 | 0.93 | 0.48 | 4 |
| $a_1^3(2)$ | EOV | 6–16 | 1.023 | 0.040 | 0.58 | 0.77 | 4 |
| $a_1^6(2)$ | EOV | 8–16 | 0.927 | 0.044 | 0.77 | 0.57 | 4 |
| $a_1(3)$ | EOV | NP | NP | NP | NP | NP | 4 |
| $b_1$ | point | NP | NP | NP | NP | NP | 4 |
| $b_1$ | EOV | 7–16 | 0.823 | 0.067 | 1.33 | 0.24 | 4 |
| $b_1$ | C | 8–16 | 1.011 | 0.096 | 1.00 | 0.42 | 4 |
| $b_1^3(1)$ | EOV | 6–16 | 0.973 | 0.044 | 0.99 | 0.43 | 4 |
| $b_1^6(1)$ | EOV | 7–16 | 0.843 | 0.026 | 0.95 | 0.46 | 4 |
| $h_1/b_1$ | EOV | 7–16 | 0.814 | 0.050 | 1.72 | 0.11 | 4 |
| $b_1^6(2)$ | EOV | 7–16 | 0.855 | 0.032 | 0.65 | 0.69 | 4 |
| $b_1(3)$ | EOV | 7–16 | 0.873 | 0.063 | 1.96 | 0.07 | 4 |

Table 4: Hadron masses $am_q = 0.025$. Notation: superscript is dimension of representation of time slice group; number of links in parenthesis; tilde (˜) state has extra $\gamma_0$; notation abbreviated when unambiguous.

| $\beta = 5.85$ ||||||||
| $am_q = 0.01$ |||| $am_q = 0.025$ ||||
| particle | range | mass | error | particle | range | mass | error |
|---|---|---|---|---|---|---|---|
| $\pi$ | 7–14 | 0.2743 | 0.0005 | $\pi$ | 9–14 | 0.4243 | 0.0008 |
| $\tilde{\pi}$ | 7–15 | 0.4385 | 0.0080 | $\tilde{\pi}$ | 6–14 | 0.5577 | 0.0048 |
| $\rho$ | 6–14 | 0.6476 | 0.0149 | $\rho$ | 6–13 | 0.7183 | 0.0056 |
| $\tilde{\rho}$ | 4–12 | 0.6258 | 0.0084 | $\tilde{\rho}$ | 4–12 | 0.7126 | 0.0040 |
| $f_0/a_0$ | 5–13 | 0.5624 | 0.0321 | $f_0/a_0$ | 6–14 | 0.8075 | 0.0314 |
| $a_1$ | 2–9 | 0.8323 | 0.0179 | $a_1$ | 4–12 | 0.9832 | 0.0284 |
| $b_1$ | 6–14 | 1.553 | 0.539 | $b_1$ | 6–13 | 1.274 | 0.213 |
| $N$ | 5–13 | 0.9501 | 0.0276 | $N$ | 5–13 | 1.060 | 0.008 |
| $N'$ | 5–13 | 0.7290 | 0.0981 | $N'$ | 5–13 | 1.184 | 0.065 |

Table 5: Quenched hadron masses at $\beta = 5.85$.

| $\beta = 5.95$ ||||||||
| $am_q = 0.01$ |||| $am_q = 0.025$ ||||
| particle | range | mass | error | particle | range | mass | error |
|---|---|---|---|---|---|---|---|
| $\pi$ | 6–14 | 0.2501 | 0.0009 | $\pi$ | 6–14 | 0.3875 | 0.0007 |
| $\tilde{\pi}$ | 8–16 | 0.3215 | 0.0044 | $\tilde{\pi}$ | 4–12 | 0.4512 | 0.0020 |
| $\rho$ | 3–11 | 0.5159 | 0.0040 | $\rho$ | 6–14 | 0.5954 | 0.0028 |
| $\tilde{\rho}$ | 2–10 | 0.5192 | 0.0042 | $\tilde{\rho}$ | 7–15 | 0.5931 | 0.0041 |
| $f_0/0$ | 8–16 | 0.4777 | 0.0541 | $f_0/0$ | 4–12 | 0.6553 | 0.0083 |
| $a_1$ | 2–10 | 0.7184 | 0.0090 | $a_1$ | 7–15 | 0.8126 | 0.0382 |
| $b_1$ | 3–11 | 0.7073 | 0.0228 | $b_1$ | 6–14 | 0.8615 | 0.0483 |
| $N$ | 7–15 | 0.7247 | 0.0285 | $N$ | 8–16 | 0.8931 | 0.0097 |
| $N'$ | 7–15 | 1.135 | 0.220 | $N'$ | 8–16 | 0.9625 | 0.115 |

Table 6: Quenched hadron masses at $\beta = 5.95$.

| | $\beta = 5.6\ am_q = 0.01$ || $\beta = 5.95\ am_q = 0.0116$ ||
| particle | mass | error | mass | error |
|---|---|---|---|---|
| $\pi$ | 0.2680 | 0.0010 | 0.2694 | 0.0009 |
| $\tilde{\pi}$ | 0.351 | 0.004 | 0.3408 | 0.0044 |
| $\rho$ | 0.518 | 0.004 | 0.5244 | 0.0040 |
| $\tilde{\rho}$ | 0.515 | 0.004 | 0.5271 | 0.0042 |
| $N$ | 0.748 | 0.004 | 0.7426 | 0.0285 |

Table 7: Comparison between quenched hadron spectrum at $\beta = 5.95$ and $am_q = 0.0116$, and spectrum of full QCD at $\beta = 5.6$ and $am_q = 0.01$.

|          | $\beta = 5.6\ am_q = 0.025$ || $\beta = 5.95\ am_q = 0.029$ ||
| particle | mass   | error  | mass   | error  |
|----------|--------|--------|--------|--------|
| $\pi$    | 0.4189 | 0.0005 | 0.4173 | 0.0008 |
| $\tilde{\pi}$ | 0.513  | 0.006  | 0.4810 | 0.0020 |
| $\rho$   | 0.637  | 0.005  | 0.6166 | 0.0028 |
| $\tilde{\rho}$ | 0.642  | 0.004  | 0.6128 | 0.0041 |
| $N$      | 0.981  | 0.008  | 0.9380 | 0.0097 |

Table 8: Comparison between quenched hadron spectrum at $\beta = 5.95$ and $am_q = 0.029$, and spectrum of full QCD at $\beta = 5.6$ and $am_q = 0.025$.

# References

[1] For a reviews of recent progress see, D. Toussaint, Nucl. Phys. **B (Proc. Suppl.) 26**, 3 (1992); A. Ukawa, Nucl. Phys. **B (Proc. Suppl.) 30**, 3 (1993).

[2] K. Bitar *et al.*, Phys. Rev. Lett. **65**, 2106 (1990); Phys. Rev. **D42**, 3794 (1990).

[3] A. Krasnitz, Phys. Rev. **D42**, 1301 (1990).

[4] K. Bitar *et al.*, Nucl. Phys. **B (Proc. Suppl.) 20**, 362 (1991); Nucl. Phys. **B (Proc. Suppl.) 26**, 259 (1992).

[5] K. Bitar *et al.*, Nucl. Phys. **B (Proc. Suppl.) 30**, 401 (1993); Phys. Rev. **D48**, 370 (1993).

[6] M. W. Hecht *et al.*, Phys. Rev. **D47**, 285 (1993).

[7] H. C. Andersen, J. Chem. Phys. **72**, 2384 (1980); S. Duane, Nucl. Phys. **B257**, 652 (1985); S. Duane and J. Kogut, Phys. Rev. Lett. **55**, 2774 (1985); S. Gottlieb, W. Liu, D. Toussaint, R. Renken and R. Sugar, Phys. Rev. **D35**, 2531 (1987).

[8] J. E. Mandula and M. C. Ogilvie, Phys. Lett. **B248**, 156 (1990).

[9] Wall sources were first introduced by the APE collaboration and are described by E. Marinari, Nucl. Phys.,**B (Proc. Suppl) 9**, 209 (1989). Our particular implementation arose through discussions with G. Kilcup.

[10] C. Liu, Nucl. Phys. **B (Proc. Suppl.) 20**, 149 (1991).

[11] A. D. Kennedy, *Intl. J. Mod. Phys.* **C3**, 1 (1992).

[12] R. Gupta, G. Guralnik, G. W. Kilcup and S. R. Sharpe, Phys. Rev. **D43**, 2003 (1991).

[13] M. F. L. Golterman, Nucl. Phys. **B273**, 663 (1986).

[14] M. F. L. Golterman and J. Smit, Nucl. Phys. **B255**, 328 (1985).

[15] For a discussion of this fitting method see D. Toussaint, in "From Actions to Answers– Proceedings of the 1989 Theoretical Advanced Summer Institute in Particle Physics," T. DeGrand and D. Toussaint, eds., (World, 1990).

[16] S. Kim and D. K. Sinclair, Argonne preprint ANL-HEP-PR-93-29, (1993).

[17] R.Altmeyer *et al.*, Nucl. Phys. **B389**, 445 (1993).

[18] M. Fukugita, H. Mino, M. Okawa, and A. Ukawa, Nucl. Phys. **B (Proc. Suppl.) 20**, 376 (1991).

[19] P. Bacilieri, *et al.*, Phys. Lett. **B214**, 115 (1988).